\documentclass[prb,aps,twocolumn,floatfix,showpacs,letter]{revtex4}
\usepackage{bm}
\usepackage{graphicx}

\begin{document}

In a recent Letter [1] we proposed an interpretation of the
experimental measurement (Refs 4-5 of [1]) of the transmission
phase shift through a quantum dot (QD) in the Kondo regime,
deduced from a double-slit A-B interferometer (ABI). Our starting
point is the 1-D single level Anderson model (SLAM) with 2
reservoirs for which we develop a scattering theory. We
distinguished between the phase shift $\delta$ of the S-matrix
responsible for the shift $\delta_{ABI}$ in the AB oscillations
($\delta_{ABI}=\delta$), and the one controlling the conductance
$G\sim \sin^{2}\delta_{G}$ (with $\delta_{G}=\delta_{\sigma}$),
and claimed the following relation holds:
$\delta_{G}=\delta_{ABI}/2$ (or equivalently
$\delta_{\sigma}=\delta/2$). The results obtained this way are in
remarkably good agreement with experimental measurements (cf.
Figs.1 and 4)

In their comment, Aharony, Entin-Wohlman, Oreg and von Delft
(AE-WOvD) [2] question the validity of our main assertion,
claiming that it fails in some exactly known limits as the
non-interacting ($U=0$) Anderson model. The main point of our
paper, however, was that the SLAM provides an incomplete
description of the experimental device. Rather the quantum dot
needs to be viewed as an artificial atom and electrons scattering
off it must satisfy the generalized Levinson theorem that
incorporates the Pauli principle in the many electron system.
Adding this physics takes us out of the strict SLAM description.
We now provide some details.

(i) First we derive Eq.(3) of JVL. The first step consists in
evaluating the retarded Green's function of one electron on the
site $0$ for the SLAM. Using exact results for the self-energy in
an interacting Fermi liquid at $T=0$, one can show that, at any
$U$,
$\mathcal{G}_{\sigma}(\mu+i\eta)=\sin\delta_{\sigma}e^{i\delta_{\sigma}}/Im\Sigma_{\sigma}(\mu+i\eta)$,
where $\delta_{\sigma}=\pi n_{0\sigma}$ and
$Im\Sigma_{\sigma}(\mu+i\eta)=-\pi(V_{L}^{2}+V_{R}^{2})\rho_{\sigma}(\mu)$.
At $U=0$, one can check that the exact Green's function of the
SLAM (cf. expression given by AE-WOvD) satisfies the latter
expression. In a second step, one derives the S-matrix at $T=0$ in
the absence of magnetic moment from
$\hat{S}_{k\sigma}^{'}=(\hat{I}-i\hat{T}^{'}_{k\sigma})$ [3]. The
elements of $\hat{T}^{'}_{k\sigma}$ are given by the r.h.s. of
Eq.2 of JVL. Incorporating the result above for
$\mathcal{G}_{\sigma}(\varepsilon_{k}+i\eta)$, one can derive
Eq.(3) of JVL leading, in the case of a symmetric QD
($V_{L}=V_{R}$), to
\begin{equation}
\label{S}
    \hat{S}_{k_F\sigma}^{'}=e^{i\delta_{\sigma}}
    \left(
    \begin{array}{cc}
      \cos{\delta_{\sigma}} & i\sin{\delta_{\sigma}}
\\
      i\sin{\delta_{\sigma}} & \cos{\delta_{\sigma}}
\\
    \end{array}
    \right).
\end{equation}
Note that the latter equation for $\hat{S}_{k_F\sigma}^{'}$
completely agrees with Eq.(2) for $\hat{S}_{k_F\sigma}$ of AE-WOvD
when $\theta$ is taken equal to $\pi/4$ as it should be for a
symmetric QD.

(ii) The expression we have just derived violates the generalized
Levinson theorem. The Levinson theorem (see Refs.16-17 of [1] and
references within) in its generalized version relates the phase
shift at zero energy $\delta(0)$ to the number of composite bound
states $N_{B,}$ formed by the incident particle and the scatterer
(as that is usual in the standard Levinson), {\it plus} an
additional number denoted by $N_{Pauli}$, equal to the number of
states excluded by the Pauli principle, i.e. $\delta(0)=\pi
(N_{B}+N_{Pauli})$. In the case of the electron scattering by an
hydrogen atom for instance, the scattering can occur in the
singlet or triplet channel. In both cases, the phase shift is
found to be $\pi$ which either comes from the existence of a bound
state for singlet scattering, or from a number of excluded states
equal to 1 for triplet case. Applying this theorem to the problem
of the scattering of a spin $\sigma$ electron off a QD, described
by an artificial atom containing a total electron number $n_0$,
one can show that: $\delta=\pi (n_{0-\sigma} + n_{0\sigma})=\pi
n_{0}$, in which $n_{0-\sigma}$ is the number of bound states
(Kondo singlet state in the Kondo regime), and $n_{0\sigma}$ is
the number of states excluded by the Pauli principle. As
announced, the expression (1) for $\hat{S}_{k_F\sigma}^{'}$
violates the generalized Levinson theorem since
$(1/2i)ln\det\hat{S}_{k_F\sigma}^{'}=\delta_{\sigma}=\pi
n_{0\sigma}$, missing the other part related to $n_{0-\sigma}$.

(iii) Our claim and we agree with the comment of AE-WOvD, is that
the 1-D SLAM with 2 reservoirs is not sufficient to capture the
whole physics contained in the experimental device. While it
captures most of the physics, it fails to account for the many
electron nature of the experimental set-up. One may try to start
with a many level Anderson model (MLAM) description of the system.
We have chosen another route and introduced minimally the missing
ingredients through an additional multiplicative phase factor
$\textrm{C}_{\sigma}$ in front of the S-matrix of the SLAM:
$\hat{S}_{k\sigma}=\textrm{C}_{\sigma} \hat{S}_{k\sigma}^{'}$. The
value of $\textrm{C}_{\sigma}$ is determined in order to guarantee
the generalized Levinson theorem. It is easy to check that
$\textrm{C}_{\sigma}=\exp^{i\delta_{-\sigma}}$ which eventually
leads to Eq.(4) of JVL for $\hat{S}_{k_F\sigma}$. By doing so, the
total occupancy of the QD as evaluated in the 1D-SLAM is directly
related to the phase shift at T=0. We believe that this is
precisely the quantity measured in the quantum interferometry.

We acknowledge clarifying correspondence with A. Aharony and Y.
Oreg. We also thank N. Andrei, G. Montambaux, P.A. Lee (for A.J.),
K. Lehur (for P.V.), P. Nozi\`eres and P. Woelfle for very useful
discussions. This work is supported by the ANR (Agence Nationale
de la Recherche) in the framework of the QuSpins Project of the
2005 PNANO Program.

M. Lavagna and P. Vitushinsky

{\it {Commissariat \`a l'Energie Atomique DRFMC/SPSMS, 17, rue des
Martyrs, 38054 Grenoble, France}}

A. Jerez

{\it {Center for Materials Theory, Rutgers University, Piscataway,
NJ 08854, US}}

[1] A. Jerez, P. Vitushinsky, and M. Lavagna, Phys. Rev. Lett.
{\bf 95}, 127203 (2005)

[2] A. Aharony, O. Entin-Wohlman, Y. Oreg, and J. von Delft, Phys.
Rev. Lett., comment (2005).

[3] The opposite sign convention for the definition of $\hat{S}$
is taken in Eq.(2) of AE-WOvD explaining the sign difference which
is found.

\end{document}